\documentstyle[aps,prl,twocolumn]{revtex}
\begin{document}
\title{Effect of annealing on electron dephasing in three-dimensional polycrystalline metals}
\author{J. J. Lin,$^{1,\ast}$ Y. L. Zhong,$^2$ and T. J. Li$^3$}
\address{$^1$Institute of Physics, National Chiao Tung University, Hsinchu 300, Taiwan}
\address{$^2$Department of Physics, National Tsing Hua University, Hsinchu 300, Taiwan}
\address{$^3$Department of Electrophysics, National Chiao Tung University, Hsinchu 300, Taiwan}
\maketitle

\begin{abstract}

We have studied the effect of thermal annealing on electron dephasing times $\tau_\phi$ in
three-dimensional polycrystalline metals. Measurements are performed on as-sputtered and annealed
AuPd and Sb thick films, using weak-localization method. In all samples, we find that $\tau_\phi$
possesses an extremely weak temperature dependence as $T\,\rightarrow\,0$. Our results show that
the effect of annealing is non-universal, and it depends strongly on the amount of disorder
quenched in the microstructures during deposition. The observed ``saturation" behavior of
$\tau_\phi$ cannot be easily explained by magnetic scattering. We suggest that the issue of
saturation can be better addressed in three-dimensional, rather than lower-dimensional,
structures.

\end{abstract}

\section{Introduction}

The electron dephasing time $\tau_\phi$ is a very important quantity that governs the mesoscopic
phenomena at low temperatures. Particularly, the behavior of the dephasing time near zero
temperature, $\tau_\phi^0 = \tau_\phi (T \rightarrow 0)$, has recently attracted vigorous
experimental \cite{Webb97,Saclay00,Lin01,Natelson01,Ovadyahu01,Bird99,Marcus99} and theoretical
\cite{Zaikin98,Alt98,Zawa99,Imry99,Wu01,Buttiker01} attention. One of the central themes of this
renewed interest is concerned with whether $\tau_\phi^0$ should reach a finite or an infinite
value as $T \rightarrow$ 0. The connection of the zero-temperature dephasing behavior with the
very fundamental condensed matter physics problems such as the validity of the Fermi-liquid
picture, the possibility of the occurrence of a quantum phase transition, and the persistent
currents in metals, etc., has been addressed. Conventionally, it is accepted that $\tau_\phi^0$
should reach an infinite value if there exist only the inelastic electron-electron and
electron-phonon scattering. However, several recent measurements performed on different mesoscopic
conductors have revealed that $\tau_\phi^0$ depends only very weakly on $T$, if at all, when $T$
is sufficiently low. There is no generally accepted process of electron--low-energy-excitation
interactions that can satisfactorily explain the ``saturation" of $\tau_\phi^0$ found in the
experiments. It should be noted that those experiments
\cite{Webb97,Lin01,Natelson01,Bird99,Marcus99} have ruled out electron heating, external microwave
noises, and very dilute magnetic impurities as the origins for the observed finite dephasing as $T
\rightarrow 0$.

To unravel the issue of electron dephasing, systematic information of $\tau_\phi^0$ over a wide
range of sample properties is very desirable. Bearing this in mind, we have in this work performed
systematic measurements of $\tau_\phi^0$ on several series of {\em as-sputtered} and subsequently
{\em annealed} AuPd and Sb {\em thick} films. The low-field magnetoresistances of the as-sputtered
samples are first measured. The samples are then annealed, and their magnetoresistances measured.
The annealing and magnetoresistance measurement procedures are repeated a few times. $\tau_\phi$
is extracted by comparing the measured magnetoresistances with the three-dimensional (3D)
weak-localization (WL) theoretical predictions \cite{Fuku81}. Generally, thermal annealing causes
a decrease in the sample resistivity, signifying a reduction in the amount of defects in the
microstructures. Controlled annealing measurements are thus crucial for testing the theoretical
models of electron dephasing invoking magnetic impurities and dynamical defects \cite{Lin87}.

\section{Experimental Method}

Thick film samples were prepared by dc sputtering deposition onto glass substrates held at room
temperature. The deposition rate was varied to tune the amount of disorder, i.e., the residual
resistivity $\rho_0$ (= $\rho$(10 K)) of the films. The AuPd films were typically 6000 $\rm\AA$
$\times$ 0.3 mm $\times$ 17 mm, while the Sb films were typically 3000 $\rm\AA$ $\times$ 0.3 mm
$\times$ 17 mm. Thermal annealing of the AuPd (Sb) films was performed in a 99.999\% pure Ar
atmosphere at moderate temperatures of $\sim$ 100$-$300$^\circ$C ($\sim$ 150$^\circ$C) for about
one half to several hours until $\rho_0$ changed by a desirable amount. The use of an extremely
high purity Ar atmosphere greatly minimized the presence of any oxygen residual gas in the
annealing. The values of the relevant parameters for our as-sputtered films are listed in
table~\ref{t.1}.

\begin{table}
\caption{Resistivities, diffusion constants, and dephasing times of the as-sputtered AuPd and Sb
thick films. The value of $\tau_\phi^0$ is extracted by least-squares fitting the measured
$\tau_\phi (T)$ to eq. ${\rm (1)}$.}

\begin{tabular}{lcccc}
Film & $\rho$(300 K) ($\mu \Omega$\,cm) & $\rho_0$ ($\mu \Omega$\,cm) & $D$(cm$^2$/s) &
$\tau_\phi^0$ ($10^{-10}$ s) \\ \hline
AuPd1e & 131 & 124 & 4.9 & 0.18 \\
AuPd4a & 509 & 467 & 1.3 & 0.85 \\
AuPd5e & 535 & 473 & 1.3 & 0.88 \\
AuPd6e & 117 & 115 & 5.3 & 0.14 \\
Sb01B & 701 & 746  & 5.8 & 0.24 \\
Sb12 & 1485 & 1645 & 2.6 & 0.51
\end{tabular}
\label{t.1}
\end{table}

We notice that the four AuPd films listed in Table \ref{t.1} were {\em newly} made from a {\em
new} Au$_{50}$Pd$_{50}$ target different from that used in our previous study \cite{Lin01}.
Moreover, a {\em different} sputtering gun and a different vacuum chamber were utilized.
Previously, we had studied $\tau_\phi$ in a series of dc sputtered AuPd thick films prepared and
measured in a different laboratory \cite{Lin01,Lin98}. By applying these new samples, we are able
to perform a close comparison study of $\tau_\phi$ in the same material prepared under different
conditions. Such a comparison is indispensable for clarifying the possible role, if at all, of
magnetic scattering on $\tau_\phi^0$. If there were any noticeable magnetic contamination during
this experiment, it is natural to expect an {\em unintentional} magnetic concentration, $n_m$,
that differs from that in our previously samples \cite{Lin01}. Consequently, a distinct value of
$\tau_\phi^0$ should be observed. On the other hand, if a similar value of $\tau_\phi^0$ is
measured, this result must then bear important information about an intrinsic material property.

In addition to the newly prepared AuPd samples, we have studied two ``aged" Sb films. The two Sb
films listed in Table \ref{t.1} were first deposited and studied two years ago in ref.
\cite{Lin00}. During the past two years, they were exposed to air all the time. One might have
naively speculated that these two samples must be heavily contaminated by (magnetic) impurities
and, thus, have a shorter $\tau_\phi$ with a much weaker $T$ dependence, compared with that
measured two years ago. To the contrary, this experiment indicates that 3D AuPd and Sb are {\em
not} as vulnerable to contamination as speculated. Our results point to an experimental fact that
suggests that the observed saturation of $\tau_\phi^0$ cannot be readily explained by magnetic
scattering.

\section{Results and Discussion}

We have measured the magnetoresistances and compared with 3D WL predictions \cite{Fuku81} to
extract the values of $\tau_\phi$. Our experimental method and data analysis procedure had been
discussed previously \cite{Lin94}. Here we emphasize that, in the limit of strong spin-orbit
scattering (which applies for both AuPd and Sb), $\tau_\phi$ is the {\em only} adjusting parameter
in the least-squares fits of the measured magnetoresistances with WL predictions. This great
advantage makes the extraction of $\tau_\phi$ highly reliable. Empirically, $\tau_\phi$ can be
written in the form
\begin{equation}
1/\tau_\phi (T) = 1/\tau_\phi^0 + 1/\tau_{\rm i} (T) \,,
\end{equation}
where $\tau_\phi^0$ dominates at the lowest measurement temperatures, and $\tau_{\rm i}$ is the
relevant inelastic scattering time which is usually important at a (few) degree(s) Kelvin and
higher. In three dimensions, electron-phonon scattering is the predominant inelastic process while
the Nyquist electron-electron scattering is negligibly small \cite{Gershen99}, i.e. $1/\tau_{\rm
i} \approx 1/\tau_{ep}$ in eq. (1). The electron-phonon scattering rate $1/\tau_{ep}$ varies as
$T^p$, with 2 $\leq p \leq$ 4.

Figure~\ref{f.1}(a) shows our measured $\tau_\phi$ as a function of temperature for four
as-sputtered AuPd films. This figure demonstrates that $\tau_\phi$ first increases with decreasing
$T$ at a few degrees Kelvin, where the electron-phonon scattering dominates the total dephasing
and $1/\tau_\phi \approx 1/\tau_{ep} \propto T^2$ in AuPd \cite{Lin98}. Below about 2 K, the
inelastic process is much less effective and a new mechanism progressively takes over, resulting
in a very weak temperature dependence of $\tau_\phi$ as $T \rightarrow 0$. To our knowledge, there
is no generally accepted process of electron--low-energy-excitation interactions that can account
for such a weak $T$ behavior. A weak temperature dependence of $\tau_\phi^0$ is also observed in
the two Sb thick films listed in table~\ref{t.1}. In fact, an (almost) absence of temperature
dependence of $\tau_\phi$ has previously been found in numerous three-dimensional polycrystalline
metals \cite{Lin01}. We notice that the values of $\tau_\phi^0$ in fig.~\ref{f.1}(a) follow {\em
closely} the scaling relation of $\tau_\phi^0 \approx 10^{-10} D^{-1}$ s established in fig. 3 of
\cite{Lin01}, where the diffusion constant $D$ is in cm$^2$/s. This result suggests that the
behavior of $\tau_\phi$ found in fig.~\ref{f.1}(a) is material intrinsic.

It should be emphasized that the weak $T$ dependence or the so-called ``saturation" of
$\tau_\phi^0$ in fig.~\ref{f.1}(a) is observed in a temperature regime where the sample resistance
varies as $-\sqrt{T}$ all the way down to our lowest measurement temperatures (fig.~\ref{f.1}(b)).
This $-\sqrt{T}$ dependence of resistance is well described by the 3D electron-electron
interaction effects \cite{Alt85}. This result indicates that the saturation of $\tau_\phi^0$ is
{\em not} caused by electron heating. A similar assertion of non-hot-electron effects has also
been reached in previous experiments \cite{Webb97,Lin01,Natelson01,Bird99,Marcus99}.

Information about the effect of annealing is crucial for clarifying the nature of magnetic
scattering and dynamical defects. Figure~\ref{f.2}(a) shows the variation of $\tau_\phi$ with
temperature for the as-prepared and subsequently annealed AuPd1e thick film. This figure clearly
indicates that $\tau_\phi$ increases with annealing. Similar behavior of increasing $\tau_\phi$
with annealing has also been found in the as-prepared and annealed AuPd6e thick film. At first
glance, this observation is easily explained. Suppose that annealing results in the rearrangement
of lattice atoms and relaxation of grain boundaries and, hence, makes the film less disordered.
Because two-level systems (TLS) are closely associated with the presence of defects in the
microstructures, their number concentration would be reduced by annealing. By assuming that
dynamical defects are effective scatterers, one can then understand fig.~\ref{f.2}(a) in terms of
a reducing TLS picture. However, our further measurements indicate that the nature of
low-temperature dephasing in real metals is not so straightforward. We find that the effect of
annealing on $\tau_\phi$ is distinctly different in strongly disordered samples (fig.~\ref{f.3}).
For the convenience of discussion, the AuPd1e and AuPd6e (AuPd4a and AuPd5e) thick films are
referred to as moderately (strongly) disordered, because they have $\rho_0 \sim$ 120 (470) $\mu
\Omega$\,cm before annealing.

Figure~\ref{f.2}(b) shows the variation of $\tau_\phi$ with $T$ for the as-prepared and annealed
Sb01B thick film. This figure clearly indicates that $\tau_\phi$ increases with annealing. Similar
effect of annealing has also been found in the Sb12 thick film. The results of figs.~\ref{f.2}(a)
and \ref{f.2}(b) suggest that an enhancement of $\tau_\phi$ by thermal annealing is common to
different {\em moderately} disordered metals. (We notice that the high resistivities in Sb films
arise from a low carrier concentration instead of a short electron mean free path \cite{Lin00}.
Our Sb thick films are thus moderately disordered.)

\section{The importance of three-dimensional structures}

One of the widely accepted explanations for the ``saturation" behavior of $\tau_\phi^0$ invokes
magnetic spin-spin scattering due to a low level contamination of the sample. This explanation has
been challenged in several recent careful experiments
\cite{Webb97,Lin01,Natelson01,Bird99,Marcus99}. However, despite this experimental situation,
there is still an insisting opinion that argues for non-zero magnetic scattering in the sample. To
completely reject such an opinion is non-trivial, because it is argued that the level of
unintentional contamination is so low that it cannot be detected by the state-of-the-art material
analysis techniques. The situation becomes more serious when reduced-dimensional systems are
involved. In the case of low-dimensional structures, surface effects due to interfaces,
substrates, and paramagnetic oxidation are non-negligible. Therefore, it is not straightforward to
ascribe the observed saturation of $\tau_\phi^0$ to either intrinsic material properties or
surface/interface effects. On the other hand, this kind of ambiguity does {\it not} occur in our
{\em three-dimensional} measurements. In fact, we believe that magnetic scattering can at most
play a subdominant role in our experiment. Our reasons are given as follows. (i) Suppose that
there is a low level of magnetic contamination in our as-sputtered films. Upon annealing, the
magnetic impurity concentration $n_m$ should be left unchanged. If the original ``saturation" in
our as-sputtered samples is caused by spin-spin scattering, one should then expect the same value
of $\tau_\phi^0$ ($\propto n_m^{-1}$) after annealing. However, we find increasing $\tau_\phi^0$
with annealing. Our result is thus in disagreement with this assumption. (ii) Our argument for a
non-magnetic origin is supported by the observation of an {\em increased} $\tau_\phi^0$ in the
aged and annealed Sb films. Since our Sb01B and Sb12 thick films were aged in air for two years,
one might have naively expected a large decrease in $\tau_\phi^0$ due to magnetic contamination.
Nevertheless, this is not the case found in fig.~\ref{f.2}(b). (iii) Moreover, if our samples do
contain an appreciable level of unintentional magnetic impurities, the contaminated concentration
$n_m$ should be basically the same in all films, because similar fabrication and measurement
procedures were involved. One should then expect a similar $\tau_\phi^0$ in all {\em as-prepared}
samples, regardless of disorder. This is certainly inconsistent with the observed scaling relation
$\tau_\phi^0 \propto D^{-1}$ discussed above. Therefore, magnetic scattering in its current form
cannot easily explain our overall results in a consistent manner.

In order to explain the widely observed saturation behavior of $\tau_\phi^0$, it has recently been
proposed that dynamical defects can be important \cite{Zawa99,Imry99}. The low-energy excitations
of the dynamical defects are usually modelled by two-level systems. We already discussed that TLS
might be partly responsible for the saturated dephasing found in our moderately disordered films.
However, it is impossible to perform a quantitative comparison of our experiment with the TLS
theories. The difficulties lie on the facts that (i) the number concentration of TLS in a
particular sample is not known, (ii) the strength of coupling between conduction electrons and a
TLS is poorly understood, and (iii) the dynamical properties of real defects (impurities, grain
boundaries, etc.) are even less clear. Experimentally, we also find other features of thermal
annealing (fig.~\ref{f.3}) that seem incompatible with a TLS picture of dephasing.

In addition to the moderately disordered samples, we have performed measurements on thick films
containing much higher levels of disorder. Surprisingly, we discover that the effect of annealing
is completely different. In the {\em strongly} disordered AuPd4a and AuPd5e thick films, we find
that annealing causes {\em negligible} effect on $\tau_\phi$. Figure~\ref{f.3} shows the variation
of $\tau_\phi$ with $T$ for the as-prepared and annealed AuPd4a thick film. This figure clearly
demonstrates that the values of $\tau_\phi$ for the as-prepared and annealed samples are
essentially the same, even though the resistance, and hence diffusion constant $D$ changed by a
factor of more than 6. The absence of an appreciable annealing effect in this case implies that,
in addition to the usual TLS addressed above, these two films contain other defects that cannot be
readily cured by thermal annealing. Such a null effect of annealing seems to suggest that, despite
a large effort in this direction, no real defects of any nature can be found to have dynamical
properties that may explain the saturation behavior of $\tau_\phi^0$. A comparison of
figs.~\ref{f.2} and \ref{f.3} strongly indicates that low-temperature dephasing is very sensitive
to the microstructures.

The observation of fig.~\ref{f.3} deserves further discussion. First, we recall that our measured
$\tau_\phi$ in the as-sputtered, strongly (and moderately) disordered films follows the scaling
relation of $\tau_\phi^0 \propto D^{-1}$ mentioned above \cite{Lin01}, implying that the result of
fig.~\ref{f.3} is material intrinsic. Secondly, in the context of magnetic scattering, Blachly and
Giordano recently found that Kondo effect is very {\em sensitive} to disorder, namely, an increase
in disorder suppresses Kondo effect \cite{ng94}. Along this line, if the original saturated
$\tau_\phi^0$ found in fig.~\ref{f.3} were due to magnetic scattering, one should argue that
thermal annealing that suppresses disorder should enhance Kondo effect. Then, a decreased
$\tau_\phi^0$ should be expected with annealing. Since our $\tau_\phi^0$ does not change even when
the sample resistivity is reduced by a factor of more than 6 by annealing, fig.~\ref{f.3} thus
cannot be easily reconciled with a magnetic scattering scenario. Besides, this picture of a
weakened Kondo effect by disorder is also incompatible with our result for the moderately
disordered films (fig.~\ref{f.2}) where an increased, instead of a decreased, $\tau_\phi^0$ with
annealing is found. Thirdly, one might argue that annealing in the presence of oxygen can lead to
the oxidation of magnetic impurities, and hence to the disappearance of Kondo resistivity
\cite{Fickett74}. We argue that this is unlikely in our case, since we had taken precautions by
annealing our films in a 99.999\% pure Ar atmosphere, where the amount of oxygen residual gas was
greatly minimized. Moreover, since we do not find a rapid increase in the $T$ dependence of
$\tau_\phi^0$ after annealing, there is thus no clear evidence of the presence of oxygen or
magnetic impurities in our films. In any case, it would be interesting to study if annealing in
oxygen would have a drastic effect on $\tau_\phi^0$.

Lastly, we discuss the advantage of using three-, instead of lower-dimensional, mesoscopic
structures for $\tau_\phi$ measurements. In 3D, the dominating inelastic process is the
electron-phonon scattering for which $\tau_{ep}$ obeys a strong $T^{-p}$ (2 $\leq p \leq$ 4)
dependence \cite{Lin98,Gershen99}. Such a temperature variation is much stronger than the
dominating $p = 2/3$ in one dimension and $p =1$ in two dimensions (which are both due to
electron-electron scattering \cite{Alt85}). Inspection of the solid lines, which are drawn
proportional to $T^{-2}$, in figs.~\ref{f.1} and \ref{f.3} reveals that our measured $\tau_\phi^0$
at 0.3 K is already $\sim$ {\em two orders of magnitude} lower than that as would be extrapolated
from the measured $\tau_{ep}$ at a few degrees Kelvin. Such a huge discrepancy is well outside any
experimental uncertainties. On the contrary, in the case of narrow wires, the dominating inelastic
dephasing time obeys a much weaker $T^{-2/3}$ law just mentioned. In this case, any discrepancy
between the measured and extrapolated values of $\tau_\phi^0$ would be less dramatic in the
attainable experimental temperature range, rendering a discrimination of the presence or absence
of a saturated $\tau_\phi^0$ less clear-cut.

\section{Conclusion}

We have studied the influence of thermal annealing on low-temperature electron dephasing in
polycrystalline AuPd and Sb thick films. We find that $\tau_\phi^0$ reveals an extremely weak
temperature dependence in both as-sputtered and annealed samples. The effect of annealing is
non-universal, depending strongly on the amount of disorder quenched in the microstructures during
deposition. The observed saturation behavior of $\tau_\phi^0$ cannot be easily explained by
magnetic-scattering in its current form. We also find that the disorder behavior of $\tau_\phi^0$
in as-prepared and annealed samples is very different. A complete theoretical explanation would
need to take the microstructures into account.

\acknowledgments The authors are grateful to M. B\"{u}ttiker, C. S. Chu, P. Mohanty, R. Rosenbaum,
J. von Delft, G. Y. Wu, and A. Zawadowski for valuable discussion. This work was supported by the
Taiwan NSC through Grant No. NSC 89-2112-M-009-054.

\begin{figure}
\caption{(a) $\tau_\phi$ as a function of temperature for four as-prepared AuPd thick films. (b)
$\triangle R/R(T) = [R(T) - R {\rm (10\,K)}]/R {\rm (10\,K)}$ as a function of $\sqrt{T}$ for
three AuPd thick films. \label{f.1}}

\vspace{6mm} \caption{$\tau_\phi$ as a function of temperature for (a) the as-prepared and
annealed AuPd1e thick film, and (b) the as-prepared and annealed Sb01B thick film. \label{f.2}}

\vspace{6mm} \caption{$\tau_\phi$ as a function of temperature for the as-prepared and annealed
AuPd4a thick film. \label{f.3}}

\end{figure}


\begin{references}

\bibitem[*]  *E-mail: jjlin@cc.nctu.edu.tw

\bibitem{Webb97} Mohanty P. {\it et al.}, Phys. Rev. Lett. {\bf 78} 3366, 1997; Webb R. A. {\it et al.},
Fortschr. Phys. {\bf 46} 779, 1998; Webb R. A. {\it et al.}, in {\it Quantum Coherence and
Decoherence}, ed. by K. Fujikawa and Y. A. Ono (Elsevier, Amsterdam, 1999).

\bibitem{Saclay00} Gougam A.B. {\it et al.}, J. Low Temp. Phys. {\bf 118} 447, 2000;
Pierre F. {\it et al.}, cond-mat/0012038.

\bibitem{Lin01} Lin J.J., and Kao L.Y., J. Phys.: Condens. Matter {\bf 13} L119, 2001.

\bibitem{Natelson01} Natelson D. {\it et al.}, Phys. Rev. Lett. {\bf 86} 1821, 2001.

\bibitem{Ovadyahu01} Ovadyahu Z., Phys. Rev. B {\bf 63} 235403, 2001.

\bibitem{Bird99} Pivin, Jr., D.P. {\it et al.}, Phys. Rev. Lett. {\bf 82} 4687, 1999.

\bibitem{Marcus99} Huibers A.G. {\it et al.}, Phys. Rev. Lett. {\bf 83} 5090, 1999.

\bibitem{Zaikin98} Golubev D.S. and Zaikin A.D., Phys. Rev. Lett. {\bf 81} 1074, 1998;
cond-mat/0103166.

\bibitem{Alt98} Aleiner I.L. {\it et al.}, Wave Random Media {\bf 9} 201, 1999; Altshuler B.L.
{\it et al}, Physica E {\bf 3} 58, 1998.

\bibitem{Zawa99} Zawadowski A., von Delft J., and Ralph D.C., Phys. Rev. Lett. {\bf 83}
2632, 1999.

\bibitem{Imry99} Imry Y., Fukuyama H., and Schwab P., Europhys. Lett. {\bf 47} 608, 1999.

\bibitem{Wu01} Wu G.Y., Phys. Rev. B, (accepted).

\bibitem{Buttiker01} B\"{u}ttiker M., cond-mat/0106149; Nagaev K.E. and B\"{u}ttiker M., cond-mat/0108243.

\bibitem{Fuku81} Fukuyama H. and Hoshino K., J. Phys. Soc. Jpn. {\bf 50} 2131, 1981.

\bibitem{Lin87} Lin J.J. and Giordano N., Phys. Rev. B {\bf 35} 1071, 1987.

\bibitem{Lin98} Zhong Y.L. and Lin J.J., Phys. Rev. Lett. {\bf 80} 588, 1998.

\bibitem{Lin00} Lin J.J., Li T.J., and Wu T.M., Phys. Rev. B {\bf 61} 3170, 2000.

\bibitem{Lin94} Wu C.Y. and Lin J.J., Phys. Rev. B {\bf 50} 385, 1994.

\bibitem{Gershen99} Gershenson M.E., Ann. Phys. (Leipzig) {\bf 8} 559, 1999; Wu C. Y. {\it et al},
Phys. Rev. B {\bf 57} 11232, 1998.

\bibitem{Alt85} Altshuler B. L. and Aronov A. G., in {\it Electron-Electron Interactions in
Disordered Systems}, ed. by A. L. Efros and M. Pollark (Elsevier, Amsterdam, 1985).

\bibitem{ng94} Blachly M. A. and Giordano N., Europhys. Lett. {\bf 27} 687, 1994;
Phys. Rev. B {\bf 51} 12537, 1995.

\bibitem{Fickett74} Fickett F. R., Mater. Sci. Eng. {\bf 14} 199, 1974.

\end{references}
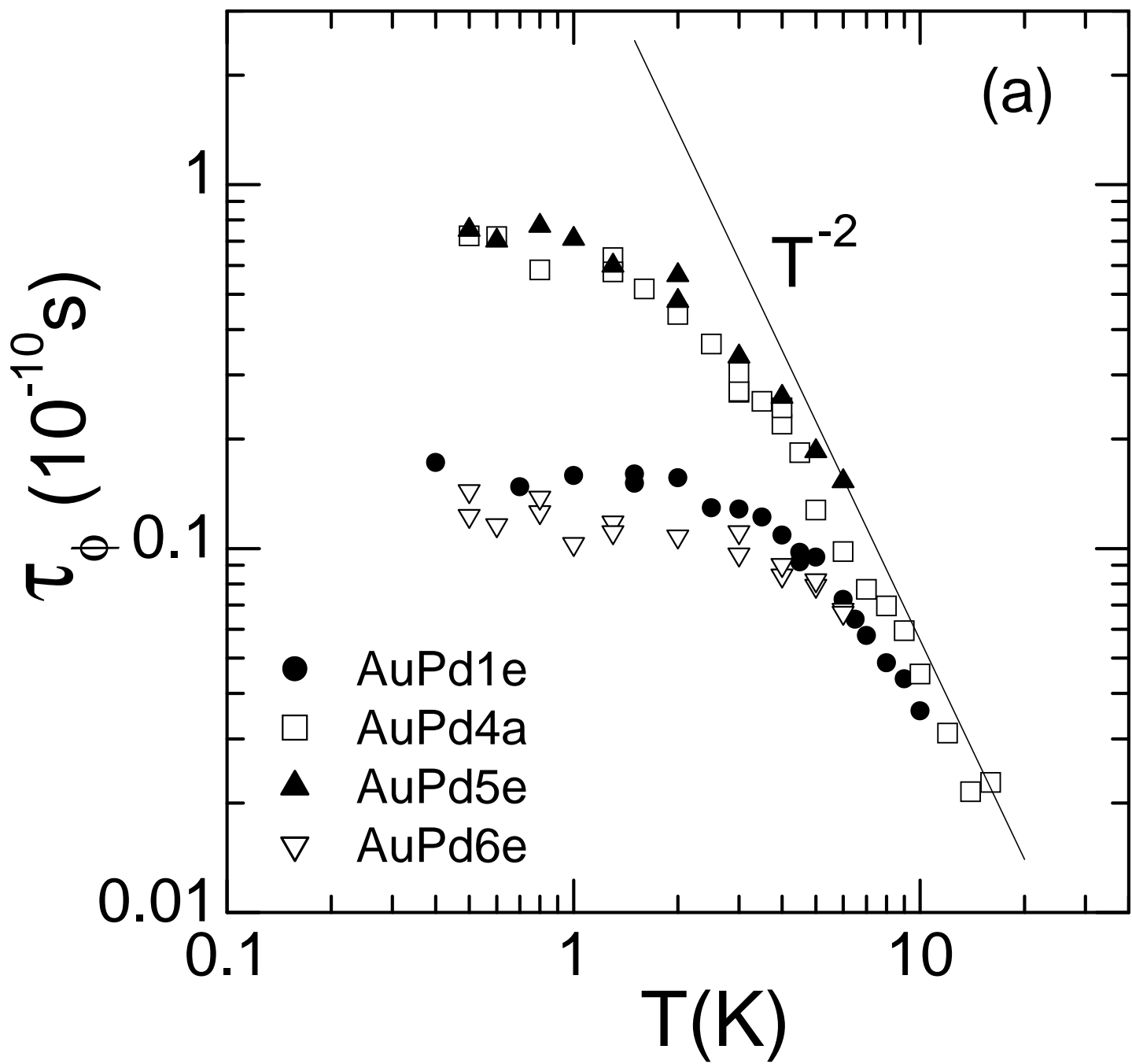
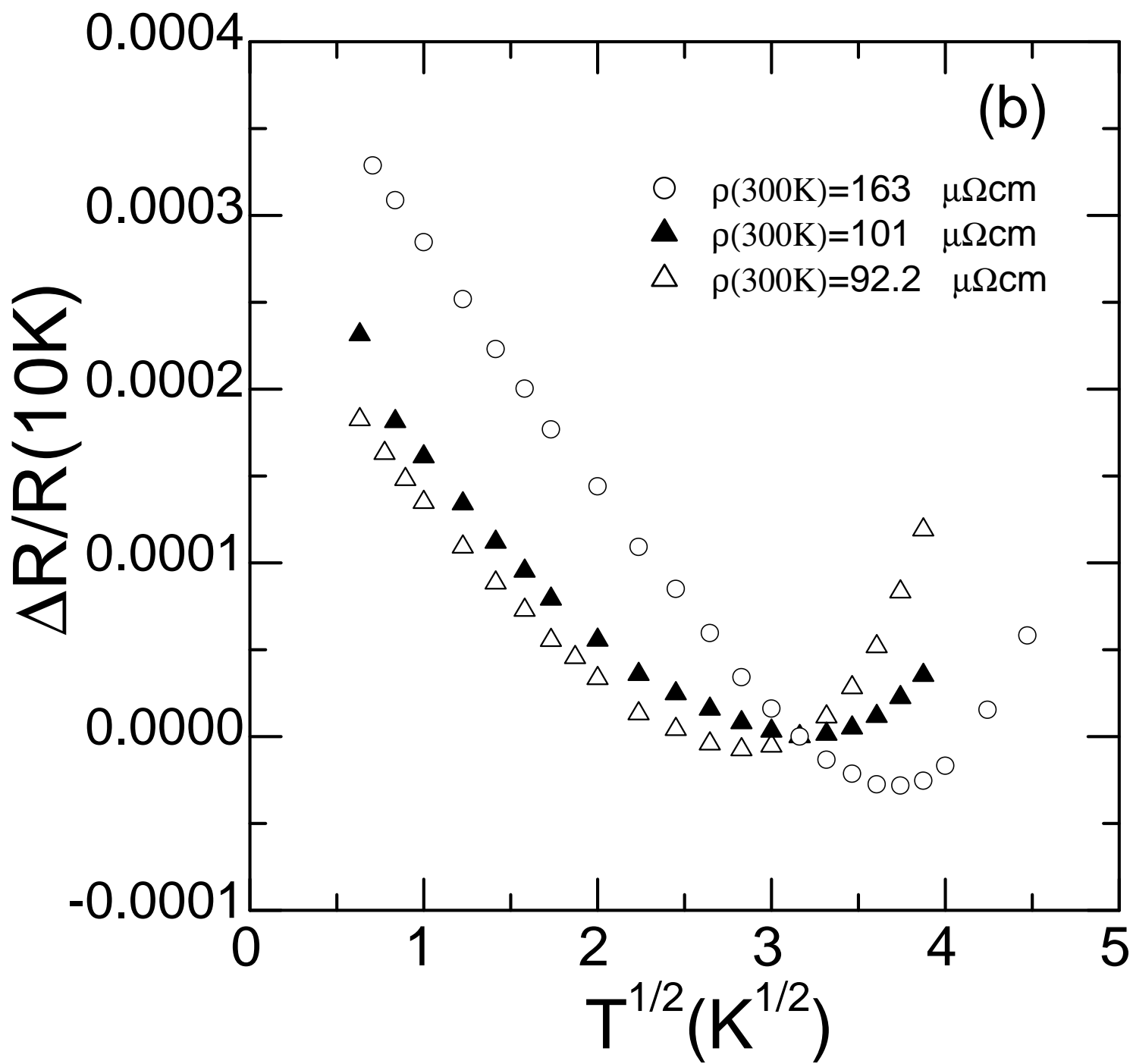
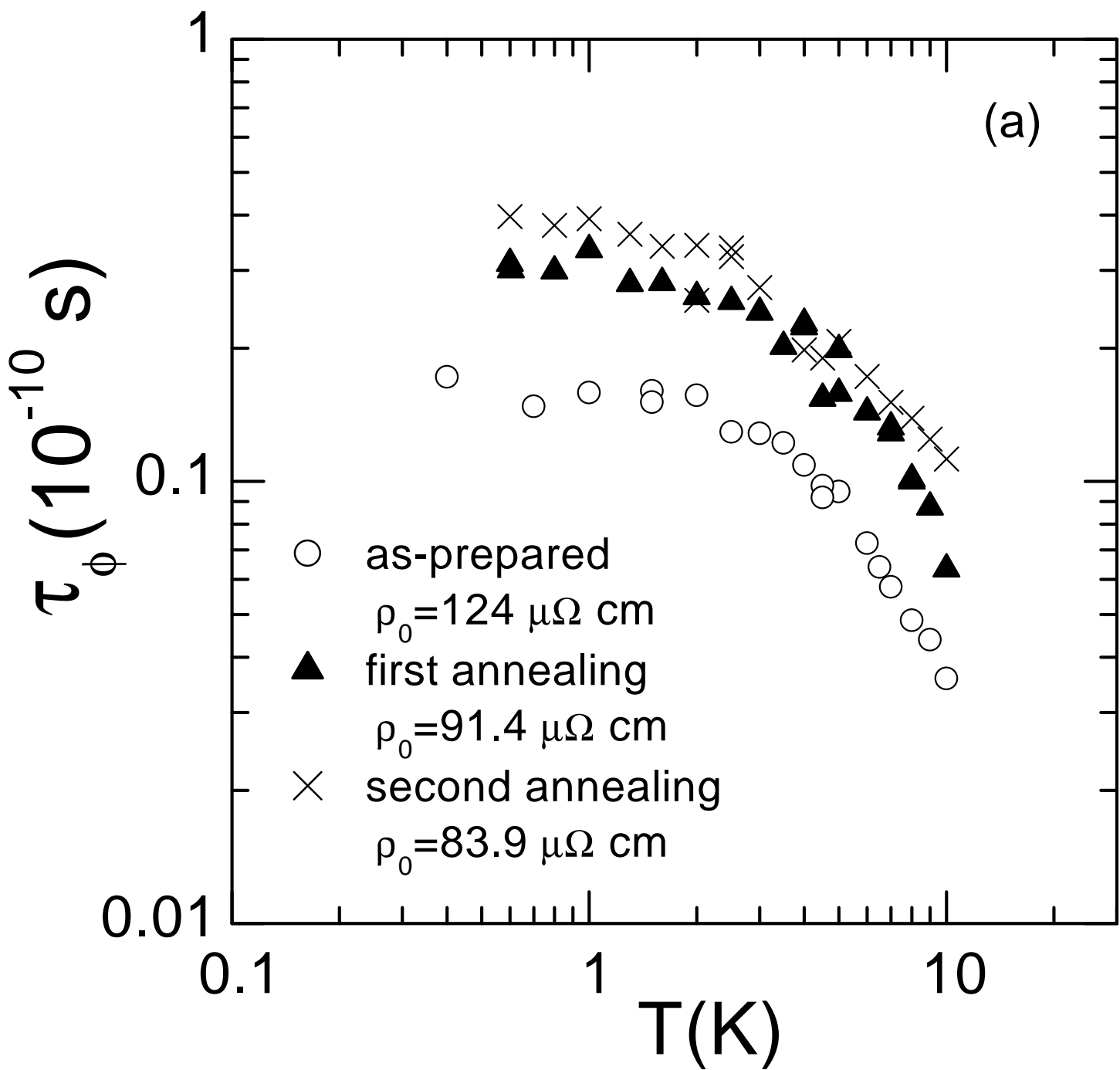
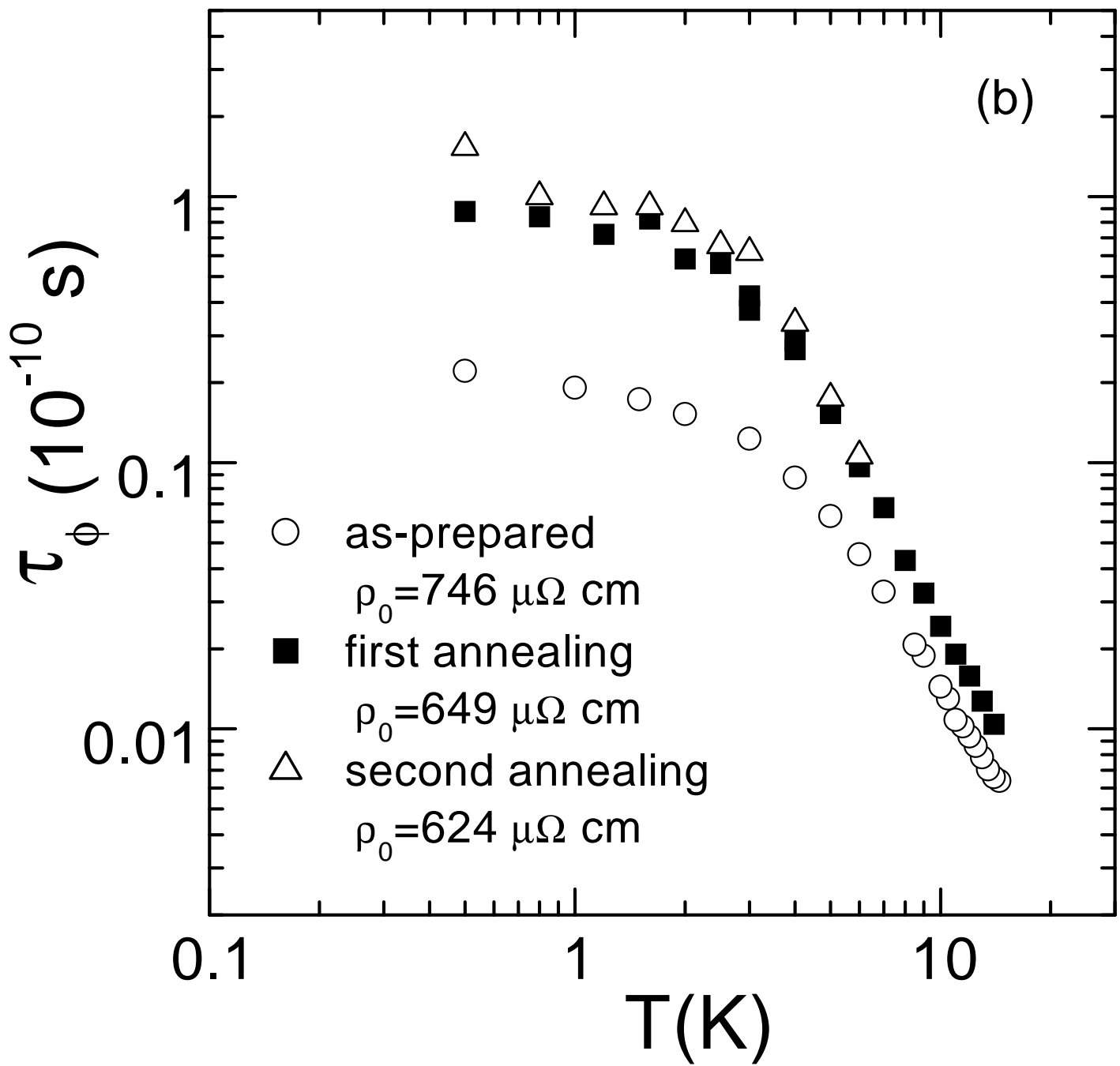
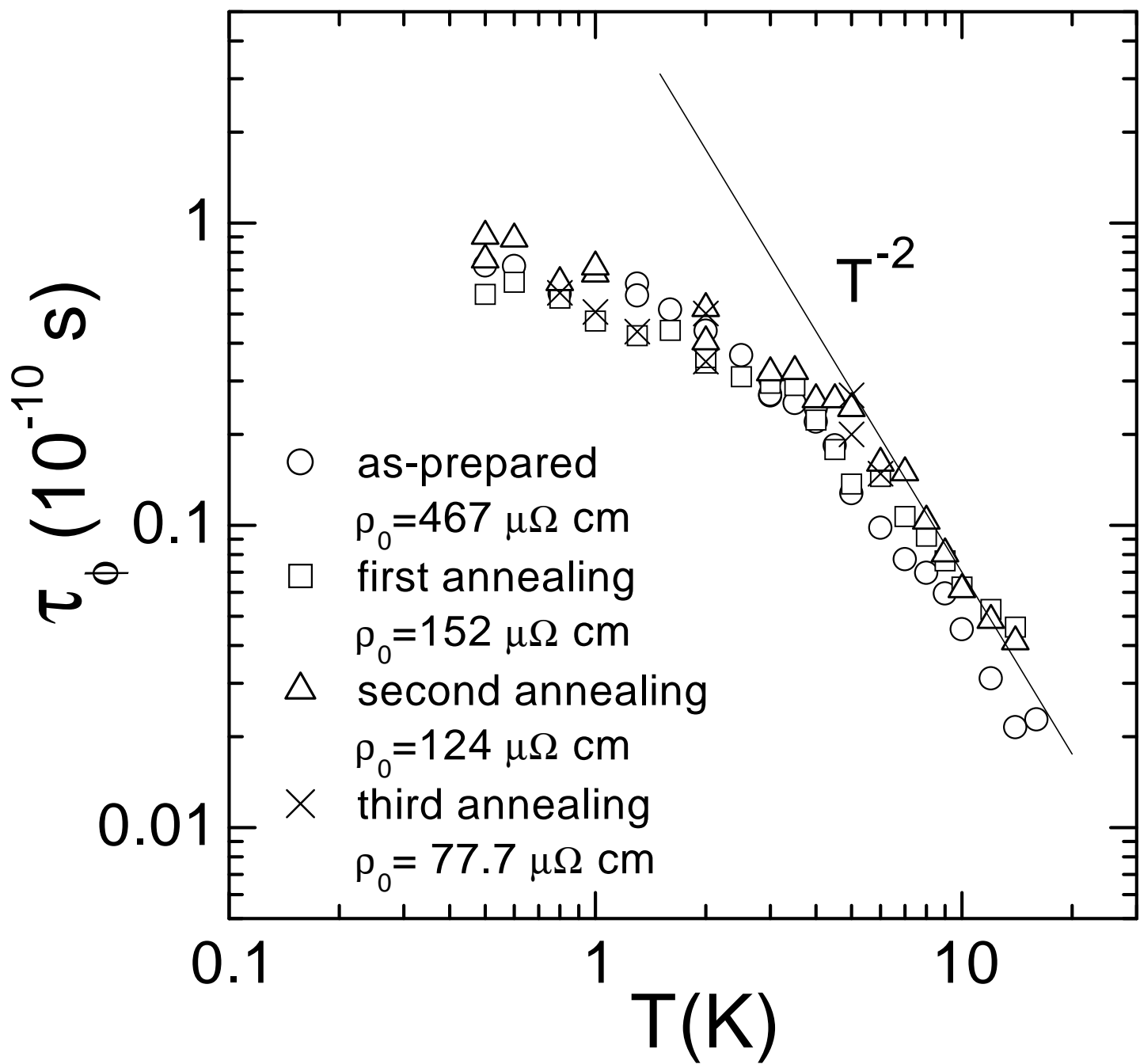
\end{document}